\documentclass[3p]{elsarticle}
\usepackage{epstopdf}
\usepackage{subfigure,subcaption,graphicx,float}
\usepackage{amsmath, amsfonts, amssymb}
\usepackage{amsthm}
\usepackage{yhmath}
\usepackage{epsf}
\usepackage{amsfonts}
\usepackage{stmaryrd}
\usepackage{amssymb}
\usepackage{leftidx}
\usepackage{xcolor}
\usepackage{mathtools}
\usepackage{placeins}
\usepackage{booktabs}
\usepackage{enumitem}
\usepackage{caption}
\usepackage{tabu}
\usepackage{multirow}
\usepackage{stackengine}
\usepackage[utf8]{inputenc}
\usepackage[english]{babel}
\usepackage{bm}
\usepackage{array}
\usepackage{multirow}
\def\bfu{{\bf u}}

\def\bfE{{\bf E}}

\def\bfI{{\bf I}}

\def\bfN{{\bf N}}

\def\bfX{{\bf X}}

\def\eps{\varepsilon}

\def\e0{\varepsilon_0}

\def\s0{\sigma_0}

\def\sts{\sigma_{\texttt{ts}}}
\def\scs{\sigma_{\texttt{cs}}}
\def\shs{\sigma_{\texttt{hs}}}
\def\de{\delta^\varepsilon}
\def\ce{c_{\texttt{e}}}

\DeclareMathAlphabet{\mathsfit}{T1}{\sfdefault}{\mddefault}{\sldefault}
\SetMathAlphabet{\mathsfit}{bold}{T1}{\sfdefault}{\bfdefault}{\sldefault}

\theoremstyle{plain}
\newtheorem{theorem}{Theorem}

\newtheorem{remark}[theorem]{Remark} %SYMBOL DEFINITIONS

\long\def\symbolfootnote[#1]#2{\begingroup%
\def\thefootnote{\fnsymbol{footnote}}\footnote[#1]{#2}\endgroup}

\begin{document}
\begin{frontmatter}

\title{Emergence of tension-compression asymmetry from a complete phase-field approach to brittle fracture \vspace{0.1cm}}

\vspace{-0.1cm}

\author{Chang Liu}
\ead{cliu723@gatech.edu}

\author{Aditya Kumar\corref{cor1}}
\ead{aditya.kumar@ce.gatech.edu}

\address{School of Civil and Environmental Engineering, Georgia Institute of Technology, Atlanta, GA 30332, USA \vspace{0.05cm}}

\cortext[cor1]{Corresponding author}

\begin{abstract}

\vspace{-0.1cm}

The classical variational approach to brittle fracture propagation does not distinguish between strain energy accumulation in tension versus compression and consequently results in physically unrealistic cracking under compression. A variety of energy splits have been proposed as a possible remedy. However, a unique energy split that can describe this asymmetry for general loading conditions has not been found. The main objective of this paper is to show that a complete phase-field theory of brittle fracture nucleation and propagation, one that accounts for the material strength at large, can naturally capture the tension-compression asymmetry without an energy split. One such theory has been recently proposed by Kumar et al. (2018). Over the past few years, several studies have shown that this theory is capable of accurately describing fracture nucleation and propagation for materials soft and hard under arbitrary monotonic loading conditions. However, a systematic study of the tension-compression asymmetry that emerges from this theory has not yet been reported. This paper does precisely that. In particular, this paper reports a comprehensive study of crack propagation in two problems, one involving a symmetric tension-compression state and the other involving larger compressive stresses at the crack tip. The results are compared with popular energy splits used in literature. The results show that, remarkably, for the second problem, only the complete theory is able to produce experimentally consistent results.

\endkeyword

\end{abstract}

\end{frontmatter}

\section{Introduction}

The mathematical reformulation of Griffith's \cite{Griffith21} original idea for fracture by Francfort and Marigo \cite{Francfort98} has led to an elegant theory of fracture propagation called the variational theory of brittle fracture. In this theory, the calculation of the fracture state is viewed as a minimization problem for the sum of the bulk strain energy and surface fracture energies with arbitrary add-cracks as one of the test fields. The variational theory can predict both when a large pre-existing crack starts to propagate and how it propagates. 
Phase-field models \cite{Bourdin00}, developed as a regularization of the variational theory to make it more amenable to numerical studies, retained the elegancy of the sharp theory and have proved to be a successful approach to simulate fracture propagation. { However, they also inherited a significant shortcoming of the variational theory -- they do not distinguish between strain energy accumulation in tension and compression and result in crack propagation under compression. Consequently, they lack the key physical property of tension-compression asymmetry required for a complete model for crack propagation.}

The popular remedial approach is to split or decompose the strain energy into a positive (opening) and negative (closing) part where only the positive energy drives the evolution of the scalar phase field variable.
The energy split approach keeps the variational nature of the classical phase-field model, { although it should be noted that the regularized phase-field equations are typically required to be solved with an alternating minimization approach, which breaks the variational link with the sharp model. However, this approach suffers from several shortcomings. First, it is not apparent how a unique split can be constructed for a linear elastic isotropic material \cite{AmorMarigoMaurini2009, miehe2010}, much less for an anisotropic material \cite{vanDijk2020orthotropic, vu2022anisotropic} or a finite elastic material \cite{borden2016phase, tang2019phase}. The idea of a tension-compression split is, by definition, one-dimensional and can be generalized to three dimensions in non-unique ways. Indeed, tens of splits have been proposed in the literature.  Second, many of the popular methods of energy splits are prone to having non-zero residual stiffness in the cracked regions \cite{LorenzisMaurini2023nucleation}. But perhaps most importantly, no single energy split has been yet shown in the literature to be able to comprehensively describe the tension-compression asymmetry in crack propagation for all loading conditions, static or dynamic \cite{StroblSeelig, zhang2022assessment, LorenzisMaurini2023nucleation}.}

An alternative and more natural solution may arise from the generalized approach of Kumar et al. \cite{KFLP18, KBFLP20} to phase-field modeling of nucleation and propagation of brittle fracture.
%A complete phase-field fracture model -- which a initiated by Kumar et al. \cite{KFLP18} may provide an elegant solution to these shortcomings of the classical phase-field model. 
{ In a nutshell, this theory generalizes the classical phase-field theory for fracture propagation by accounting for multiaxial crack nucleation in general through the material's strength surface, while keeping undisturbed the ability of the classical phase-field regularization to model crack propagation according to Griffith's fracture postulate.} A string of recent works \cite{KRLP18, KBFLP20, KLP20, KLP21, KRLP22, KLDLP23}  have provided a wide range of validation results for a broad spectrum of materials (silicone, titania, graphite, polyurethane, PMMA, alumina, natural rubber, glass, rock), specimen geometries (with large and small pre-existing cracks, V notches, U notches, and smooth boundaries), and loading conditions suggesting that this theory may indeed provide a complete framework for elastic brittle materials. { In the classical phase-field models, incorporation of tension-compression asymmetry in crack propagation results in a multiaxial strength surface, albeit one in which the compressive strength of the material is the same as the tensile strength. Since the Kumar et al. model accounts for the correct multiaxial strength surface of the material in the stress form in the governing equations, it was conjectured in \cite{KBFLP20}, without accompanying results, that tension-compression asymmetry in crack propagation would emerge from this model \cite{KBFLP20}.

% has been presented before in previous works for crack nucleation in the absence of pre-existing cracks. The theory was able to provide qualitative and quantitative agreement with the experiments for the indentation test \cite{KRLP22} and Brazilian fracture test \cite{KLDLP23} -- problems that both involve large compressive stresses.
The purpose of this paper is to present results on two prominent problems in the literature using the new model which give direct evidence of tension-compression asymmetry. The first example is the study of mode II fracture of a plate under a quasi-static simple shear loading. This is a benchmark problem for tension-compression asymmetry studied in many previous works \cite{Bourdin08, miehe2010}. It involves a symmetric tension-compression state in front of the crack. All energy split methods proposed in the literature can predict the correct crack path for this problem. The second example is the study of crack propagation from two inclined cracks in a plate under uniaxial compression. The rationale for our choice to study this problem is threefold. First, the analysis of this problem with the phase-field method utilizing the conventional energy splits has proved to be technically challenging \cite{zhang2017wing, BryantSun2018, wick2022}. In fact, several new energy splits have been proposed to analyze this problem \cite{wick2022}. It has been claimed in the literature that the experimental crack path can only be obtained if the material has a significant disparity in its mode I and mode II fracture toughness \cite{zhang2017wing, BryantSun2018, SteinkeKaliske2022}. We investigate here whether this is an artifact of the choice of energy split. Second, this problem is of broad interest in rock mechanics and has been extensively studied experimentally and theoretically for many decades \cite{zhou2021review, liu2022review}, indicating its challenging nature. Third, it involves an asymmetric tension-compression state at the crack tip with significantly larger compressive stresses. 
% However, it remains understudied whether the theory can describe the asymmetry under general loading conditions and whether it depends on any model parameters.  It is also understudied whether it performs better than the energy split methods for problems involving large pre-existing cracks. The main purpose of this paper is to address these questions. We do so by studying two prominent examples of fracture propagation in which the correct modeling of asymmetry plays an important role.
}
We begin in Section 2 by summarizing the general fracture theory of Kumar et al. (2018, 2022). We then present in Sections 3 and 4 a comprehensive study of the two examples. Lastly, some final comments are recorded in Section 5.

%Incorporation of tension-compression asymmetry is  Remains hugely important for materials undergoing nucleation of fracture under shear and other non-uniaxial tension stress states. The second example studies one such problem in detail and show that contrary to previous claims, wing crack nucleation can be explained for an isotropic brittle material.

\section{The complete phase-field theory of brittle fracture of Kumar et al. (2018, 2020)}\label{Sec: Fracture}

Consider a structure made of an isotropic linear elastic brittle material occupying an open bounded domain $\mathrm{\Omega}\subset \mathbb{R}^3$, with boundary $\partial\mathrm{\Omega}$ in its undeformed and stress-free configuration at time $t=0$. At a later time $t \in (0, T]$, due to an externally applied displacement $\overline{\bfu}(\bfX, t)$ on a part $\partial\mathrm{\Omega}_\mathcal{D}$ of the boundary and a traction $\overline{\textbf{t}}(\bfX,t)$ on the complementary part $\partial\mathrm{\Omega}_\mathcal{N}=\partial\mathrm{\Omega}\setminus \partial\mathrm{\Omega}_\mathcal{D}$, the structure experiences a deformation field $\bfu(\bfX,t)$. We write the infinitesimal strain tensor as
\begin{equation*}
\bfE(\bfu)=\dfrac{1}{2}(\nabla\bfu+\nabla\bfu^T).
\end{equation*}
Non-interpenetration constraint implies that ${\rm det}(\bfI+\nabla\bfu)>0$.  In response to the externally applied mechanical stimuli, cracks can also nucleate and propagate in the structure. Those are described in a regularized way by the phase field
\begin{equation*}
v=v(\bfX,t)
\end{equation*}
taking values in $[0,1]$. Precisely, $v=1$ identifies regions of the sound material, whereas $v<1$ identifies regions of the material that have been fractured.

\subsection{Constitutive behavior of the material}

For isotropic brittle materials, the mechanical behavior is assumed to be completely characterized by three intrinsic properties of the material: (\emph{i}) elasticity, (\emph{ii}) strength, and (\emph{iii}) critical energy release rate.

\paragraph{Elasticity} The elastic behavior for an isotropic linear elastic material is characterized by the stored-energy function
\begin{equation}
	W(\bfE(\bfu)) =\mu\bfE\cdot\bfE+\dfrac{\lambda}{2}({\rm tr}\,\bfE)^2,\label{W-mu}
\end{equation}
where $\mu>0$ and $\lambda>-2/3\mu$ are the Lam\'e constants. Recall the basic relations $\mu=E/(2(1+\nu))$ and $\lambda=E\nu/((1+\nu)(1-2\nu))$, where $E$ is the Young's modulus and $\nu$ is the Poisson's ratio. The stress-strain relation is given by
\begin{equation*}
	\boldsymbol{\sigma}(\bfX,t)=\dfrac{E}{1+\nu}\bfE+\dfrac{E\,\nu}{(1+\nu)(1-2\nu)}({\rm tr}\,\bfE)\bfI%\label{S-E-E}
\end{equation*}

\paragraph{Strength} When a macroscopic piece of an elastic brittle material is subjected to an arbitrary but uniform state of stress $\boldsymbol{\sigma}$, fracture nucleates at a critical value of the applied stress. The set of all such stresses defines a surface in the stress space -- called a strength surface
\begin{equation}
\mathcal{F}(\boldsymbol{\sigma})=0,\label{SSurf-0}
\end{equation}
where $\boldsymbol{\sigma}$ stands for the Cauchy stress tensor.
A popular, but not general, choice for the strength surface of the material is the Drucker-Prager strength surface
\begin{equation}
	\mathcal{F}(\boldsymbol{\sigma})=\sqrt{J_2}+\gamma_1 I_1+\gamma_0=0\qquad {\rm with}\qquad \left\{\hspace{-0.1cm}\begin{array}{l}\gamma_0=-\dfrac{2\sigma_{\texttt{cs}}\sigma_{\texttt{ts}}}
		{\sqrt{3}\left(\sigma_{\texttt{cs}}+\sigma_{\texttt{ts}}\right)}\vspace{0.2cm}\\
		\gamma_1=\dfrac{\sigma_{\texttt{cs}}-\sigma_{\texttt{ts}}}
		{\sqrt{3}\left(\sigma_{\texttt{cs}}+\sigma_{\texttt{ts}}\right)}\end{array}\right. ,\label{DP}
\end{equation}
where
\begin{equation}
	I_1={\rm tr}\,\boldsymbol{\sigma}\qquad {\rm and}\qquad  J_2=\dfrac{1}{2}{\rm tr}\,\boldsymbol{\sigma}^2_{D}\qquad {\rm with}\quad \boldsymbol{\sigma}_{D}=\boldsymbol{\sigma}-\dfrac{1}{3}({\rm tr}\,\boldsymbol{\sigma})\bfI\label{T-invariants}
\end{equation}
stand for two of the standard invariants of the stress tensor $\boldsymbol{\sigma}$, while the constants $\sigma_{\texttt{ts}}>0$ and $\sigma_{\texttt{cs}}>0$ denote the uniaxial tensile and compressive strengths of the material. 

{
\begin{remark}{\rm The two-material-parameter Drucker-Prager strength surface (1952) is arguably the simplest model that has proven capable of describing reasonably well the strength of many nominally brittle materials and has been used extensively in previous phase-field studies \cite{KBFLP20, KLP20, KRLP22, KLDLP23}. However, for many materials, other strength surfaces, such as the Mohr-Coulomb surface, may describe the experimental results better and can be used instead. Moreover, we emphasize that the strength surface only defines the crack nucleation under uniform stress states. More precisely, it defines when the phase field $v$ ceases to be 1 under uniform stress states. Under non-uniform stress states, the violation of the strength surface is not a sufficient condition for crack nucleation and requires the contribution of the third intrinsic property: critical energy release rate.
}
\end{remark}

}

%
%%%%%%%%%%%%%%%%%%%%%%%%%%%%%%%%%%%%%%%%%%%%%%%%%%%%%%%%%%%%%%%%%%%%%%%%%%%%%%
% \begin{figure}[t!]
% 	\subfigure[]{
% 		\label{fig:2a}
% 		\begin{minipage}[]{0.5\textwidth}
% 			\centering \includegraphics[width=2.3in]{Fig2a.eps}
% 			\vspace{0.2cm}
% 	\end{minipage}}
% 	\subfigure[]{
% 		\label{fig:2b}
% 		\begin{minipage}[]{0.5\textwidth}
% 			\centering \includegraphics[width=2.4in]{Fig2b.eps}
% 			\vspace{0.2cm}
% 	\end{minipage}}
% 	\caption{(a) Comparison between the Drucker-Prager strength surface (\ref{DP}), with $\sigma_{\texttt{ts}}=100$ MPa and $\sigma_{\texttt{cs}}=1232$ MPa, and strength experimental data for titania (Ely, 1972); the results are plotted for the principal stress $\sigma_2$ in terms of the principal stress $\sigma_1$ and correspond to the case when $\sigma_3=0$. (b) Plot of the Drucker-Prager strength surface (\ref{DP}), with $\sigma_{\texttt{ts}}=100$ MPa and $\sigma_{\texttt{cs}}=1232$ MPa, in the space of all three principal stresses $(\sigma_1,\sigma_2,\sigma_3)$.}\label{Fig2}
% \end{figure}
%%%%%%%%%%%%%%%%%%%%%%%%%%%%%%%%%%%%%%%%%%%%%%%%%%%%%%%%%%%%%%%%%%%%%%%%%%%%%%
%

\paragraph{Critical energy release rate} The critical energy release rate (or intrinsic fracture toughness), $G_c$ describes the total energy expended in creating unit fracture surface area. It captures the resistance to the growth of a pre-existing crack. For an isotropic brittle material, $G_c$ is a scalar material constant.

\subsection{The governing equations of deformation and fracture}

According to the theory of Kumar et al. (2018), the displacement field $\bfu_k(\bfX)=\bfu(\bfX,t_k)$ and phase field $v_k(\bfX)=v(\bfX,t_k)$ at any material point $\bfX\in\overline{\mathrm{\Omega}}$ and discrete time $t_k\in\{0=t_0,t_1,...,t_m,$ $t_{m+1},...,$ $t_M=T\}$ are determined by the system of coupled partial differential equations (PDEs)
\begin{equation}
\left\{\begin{array}{ll}
 {\rm Div}\left[v_{k}^2\dfrac{\partial W}{\partial \bfE}(\bfE(\bfu_{k}))\right]={\bf0},\quad \bfX\in\mathrm{\Omega},\\[10pt]
\bfu_{k}=\overline{\bfu}(\bfX,t_{k}),\quad \bfX\in\partial  \mathrm{\Omega}_{\mathcal{D}},\\[10pt]
 \left[v_{k}^2\dfrac{\partial W}{\partial \bfE}(\bfE(\bfu_{k}))\right]\bfN=\overline{\textbf{t}}(\bfX,t_{k}),\quad \bfX\in\partial \mathrm{\Omega}_{\mathcal{N}}\end{array}\right. \label{BVP-u-theory}
\end{equation}
and
\begin{equation}
\left\{\begin{array}{l}
\dfrac{3}{4} \varepsilon \, \de \,  G_c \triangle v_{k}=2 v_{k} W(\bfE(\bfu_{k}))-c_\texttt{e}(\bfX,t_{k})- \dfrac{3}{8}  \dfrac{\de \, G_c}{\varepsilon},
 \mbox{if } v_{k}(\bfX)< v_{k-1}(\bfX),\quad \bfX\in\mathrm{\Omega} \\[10pt]
\dfrac{3}{4} \varepsilon \, \de \,  G_c \triangle v_{k}\geq2 v_{k} W(\bfE(\bfu_{k}))-c_\texttt{e}(\bfX,t_{k})- \dfrac{3}{8} \dfrac{\de \, G_c}{\varepsilon},
 \mbox{if } v_{k}(\bfX)=1\; \mbox{ or }\; v_{k}(\bfX)= v_{k-1}(\bfX)>0,\quad \bfX\in\mathrm{\Omega} \\[10pt]
v_{k}(\bfX)=0,\quad \mbox{ if } v_{k-1}(\bfX)=0,\quad \bfX\in\mathrm{\Omega}
\\[10pt]
\nabla v_{k}\cdot\bfN=0,\quad \bfX\in \partial\mathrm{\Omega}
   \end{array}\right. \label{BVP-v-theory}
\end{equation}
with $\bfu(\bfX,0)\equiv\textbf{0}$ and $v(\bfX,0)\equiv1$, where $\nabla\bfu_k(\bfX)=\nabla\bfu(\bfX,t_k)$, $\nabla v_k(\bfX)=\nabla v(\bfX,t_k)$, $\triangle v_k(\bfX)= \triangle v(\bfX,$ $t_k)$, and where $\varepsilon>0$ is a regularization or localization length and $c_\texttt{e}(\bfX,t)$ is a driving force containing information about material's strength and $\de$ is a non-negative coefficient whose specific constitutive prescription depends in turn on the particular form of $c_\texttt{e}(\bfX,t)$ and the value of material parameters. The specific constitutive prescription for $\ce$ depends on the particular form of the strength surface and is spelled out below for the case of Drucker-Prager strength surfaces.

{
\begin{remark}{\rm The inequalities in (\ref{BVP-v-theory}) describe the two constraints on the phase field, namely that its value remains between 0 and 1 and that it decreases monotonically in time. The second constraint is due to the fact that fracture is an irreversible process in most cases. {  From an implementation point of view, we make use
of a penalty method to enforce both constraints.} The localization length $\varepsilon$ in (\ref{BVP-v-theory}) is purely a regularization parameter that is set to be smaller than the smallest characteristic length scale in the structural problem at hand. In practice, it is also set to be no larger than the material's characteristic fracture length scale.
}
\end{remark}
}

{ 
\begin{remark}{\rm The computational implementation of the equations (\ref{BVP-u-theory})--(\ref{BVP-v-theory}) differs from the implementation of the classical variational model only through the presence of the term $\ce$ on the right-hand side of the PDE for the phase-field (\ref{BVP-v-theory}). In absence of the $\ce$ term in (\ref{BVP-v-theory}), the equations (\ref{BVP-u-theory})-(\ref{BVP-v-theory}) represent the classical AT$_1$ phase-field model \cite{pham2011gradient, Tanne18}.
Accordingly, after discretizing these equations in space with finite elements, they can be solved iteratively in a staggered manner, much like the classical model. More details on the numerical implementation can be found in Kumar et al. \cite{KFLP18}. A FEniCS implementation of the numerical scheme is available in GitHub\footnote{https://github.com/farhadkama/FEniCSx\_Kamarei\_Kumar\_Lopez-Pamies}.

}
\end{remark}

{
\begin{remark}{\rm It has been noted recently by Larsen et al. \cite{larsen2024} that the equations (\ref{BVP-u-theory})--(\ref{BVP-v-theory}), which are not Euler-Lagrange equations of a variational principle,  can still be recast in a variational theory. Specifically, they have shown that the solution pair $({\bfu}, v)$ for the governing PDEs correspond to the fields that minimize separately two different functionals: (i) a deformation energy functional and (ii) a fracture
functional. They have further shown that the classical phase-field models, implemented with the alternating minimization approach, are described by the same type of variational principle. The difference between the two formulations is only in terms of the choice of the fracture functional. 
}
\end{remark}
}

}

{
\begin{remark}{\rm The main ideas behind the proposed formulation by Kumar et al. \cite{KFLP18, KBFLP20} to account for material strength can be described as follows. The first idea is that a three-dimensional description of nucleation of fracture in bulk under uniform stresses requires accounting for more than just the uniaxial tensile strength of the material. Generally, strength should be described by a scalar function of the stress tensor, called strength surface, as defined in Section 2.1. The second idea is that the strength surface, in general, can not be rewritten in terms of an energy threshold. This is the basis for the introduction of a stress-type term $\ce$ in the governing equation (\ref{BVP-v-theory}). Refer to Sections 2 and 3 of Kumar et al. \cite{KBFLP20} for a more detailed discussion. { For a thermodynamic interpretation of the theory, refer to Kumar et al. \cite{KRLP18}.}
}
\end{remark}
}

{
\begin{remark}{\rm 
Accounting for the entire strength surface naturally introduces an asymmetry in crack nucleation in tension and compression since the uniaxial compressive strength is typically larger than the uniaxial tensile strength. Moreover, the primary nature of stress, instead of energy, as the driving force for crack nucleation and propagation under large compressive forces has also been recognized with the classical energetic phase-field formulations where a stress split is employed to obtain experimentally consistent results \cite{wu2017, StroblSeelig, wick2022}.
}
\end{remark}
}

\begin{remark}{\rm The solution of the equations (\ref{BVP-u-theory})--(\ref{BVP-v-theory}) may be suspect to material interpenetration under compression. In this work, we assume geometrical linearity and only check \emph{a posteriori} that the non-interpenetration constraint is satisfied.
}
\end{remark}

\subsection{The model for the external driving force $c_{\texttt{e}}$}

The external driving force is a manifestation of the presence of the inherent defects in the material, that is, its strength surface at large. Hence, the information about the strength surface enters the governing PDEs through $\ce$. A blueprint for constructing $\ce$ was outlined in \cite{KBFLP20} for any strength surface. The construction process specifies that $\ce$ takes the same functional form as the strength surface, in this case, the Drucker-Prager strength surface (\ref{DP}), but with $\varepsilon$-dependent coefficients. Precisely, it is defined as
\begin{align}
c_{\texttt{e}}(\bfX,t)=\widehat{c}_{\texttt{e}}(I_1,J_2;\varepsilon)
=\beta_2^\varepsilon\sqrt{J_2}+\beta_1^\varepsilon I_1,\label{cehat-2020}
\end{align}
where
\begin{equation}
\left\{\begin{array}{l}
\beta^\varepsilon_1=-\dfrac{1}{\shs}\delta^\varepsilon\dfrac{G_c}{8\varepsilon}+\dfrac{2\mathcal{W}_{\texttt{hs}}}{3\shs}\vspace{0.2cm}\\
\beta^\varepsilon_2=-\dfrac{\sqrt{3}(3\shs-\sts)}{\shs\sts}\delta^\varepsilon\dfrac{G_c}{8\varepsilon}-
\dfrac{2\mathcal{W}_{\texttt{hs}}}{\sqrt{3}\shs}+\dfrac{2\sqrt{3}\mathcal{W}_{\texttt{ts}}}{\sts}\end{array}\right. , \label{betas}
\end{equation}
with
$$\mathcal{W}_{\texttt{ts}}= \dfrac{\sts^2}{2 E}, \quad \mathcal{W}_{\texttt{hs}}= \dfrac{\shs^2}{2 \kappa}, \quad \shs=\dfrac{2 \sts  \scs} { 3 (\scs - \sts)}, \quad \kappa= \dfrac{E}{3 (1 - 2 \nu)}.$$
$I_1$ and $J_2$ stand for the invariants (\ref{T-invariants}) of the Cauchy stress
\begin{equation*}
\boldsymbol{\sigma}(\bfX,t)=v^2\dfrac{\partial W}{\partial \bfE}(\bfE(\bfu))
\end{equation*}
and, hence, read as
\begin{equation*}
I_1=(3\lambda+2\mu) v^2 {\rm tr}\,\bfE(\bfu)\quad {\rm and}\quad J_2=2\mu^2 v^4 {\rm tr}\,\bfE^2_D(\bfu)
\end{equation*}
with $\bfE_D(\bfu)=\bfE(\bfu)-1/3\left({\rm tr}\,\bfE(\bfu)\right)\bfI$ in terms of the displacement field $\bfu$ and phase field $v$, and where $\delta^\varepsilon$ is a unitless $\varepsilon$-dependent coefficient. The value of $\delta^\varepsilon$ needs to be calibrated to retain Griffith-fracture propagation with the equations (\ref{BVP-u-theory})--(\ref{BVP-v-theory}). An approximate analytical formula for $\de$, recently provided in Kamarei et al. \cite{KKLP24}, is 
\begin{equation}
\delta^\varepsilon=\left(1+\dfrac{3}{8}\dfrac{h}{\varepsilon}\right)^{-2}\left(\dfrac{\sts+(1+2\sqrt{3})\,\shs}{(8+3\sqrt{3})\,\shs}\right)\dfrac{3  G_c}{16 \mathcal{W}_{\texttt{ts}} \varepsilon}+\left(1+\dfrac{3}{8}\dfrac{h}{\varepsilon}\right)^{-1}\dfrac{2}{5}.
\label{delta-eps-final-h}
\end{equation}
where $h$ is the mesh size.

% Currently, the simplest procedure to calibrate $\de$ is to solve, for a given set of material constants $E$, $\nu$, $G_c$, $\sigma_{\texttt{ts}}$, $\sigma_{\texttt{cs}}$, and a given finite localization length $\varepsilon$, a single boundary-value problem of choice for which the nucleation from a large pre-existing crack can be determined analytically and then adjust $\de$ so that the phase-field theory matches the analytical solution. An analytical formula for $\de$ is under development.

%
\begin{figure}[h!]
	\centering
	\includegraphics[width=3.in]{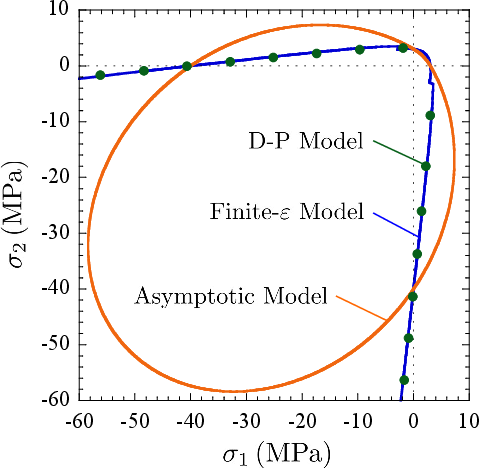}
	\caption{The strength surfaces in the ($\sigma_1$, $\sigma_2$)-space for stress states with $\sigma_3=0$ for gypsum (see Section 4) and localization length $\varepsilon=1$ mm, as predicted by the driving force introduced in Kumar et al. (2020), named Asymptotic Model, and the driving force introduced in Kumar et al. (2022), named Finite-$\varepsilon$ Model. For direct comparison, the Drucker–Prager strength surface (6) is also included.}\label{Fig1}
\end{figure}

A correction to the form for $\ce$ was presented in Kumar et al. \cite{KRLP22} to improve the description of the compressive part of the strength surface from the governing PDEs (\ref{BVP-u-theory})--(\ref{BVP-v-theory}) for large values of localization lengths $\varepsilon$. Specifically, this form of $\ce$ is defined as
\begin{align}
c_{\texttt{e}}(\bfX,t)=\widehat{c}_{\texttt{e}}(I_1,J_2;\varepsilon)
=\beta_2^\varepsilon\sqrt{J_2}+\beta_1^\varepsilon I_1+
\dfrac{1}{v^3}\left(1-\dfrac{\sqrt{I_1^2}}{I_1}\right)\left(\dfrac{J_2 (1+\nu)}{E}+\dfrac{I_1^2 (1-2 \nu)}{6 E}\right),\label{cehat-2022}
\end{align}
where $\beta_1, \beta_2$ are given by equations (\ref{betas}).
The last term is the correction term that is zero when $I_1>0$ and non-zero when $I_1<0$. In the limit of $\varepsilon \searrow 0$, the correction term is negligible as it is $O(\varepsilon^0)$ and the original model (\ref{cehat-2020}) is asymptotically obtained.
We refer to the form (\ref{cehat-2020}) for $\ce$ as the `Asymptotic Model' and the form (\ref{cehat-2022}) as the `Finite-$\varepsilon$ Model' in the text below. A direct comparison between the strength surface implied from the two models is shown in Figure \ref{Fig1}. The results are shown for gypsum studied in Section 4 with $\varepsilon=1$ mm. The figure also includes the Drucker-Prager strength surface \ref{DP} for comparison. It can be clearly observed that the Finite-$\varepsilon$ Model provides a better approximation of Drucker-Prager strength surface than the Asymptotic Model for this value of $\varepsilon$. { We finally note that a similar construction process for $\ce$ can be adopted for any other strength surface. }

\section{The simple shear problem} \label{Sec: Simpleshear}

\begin{figure}[h!]
	\centering
	\includegraphics[width=3.5in]{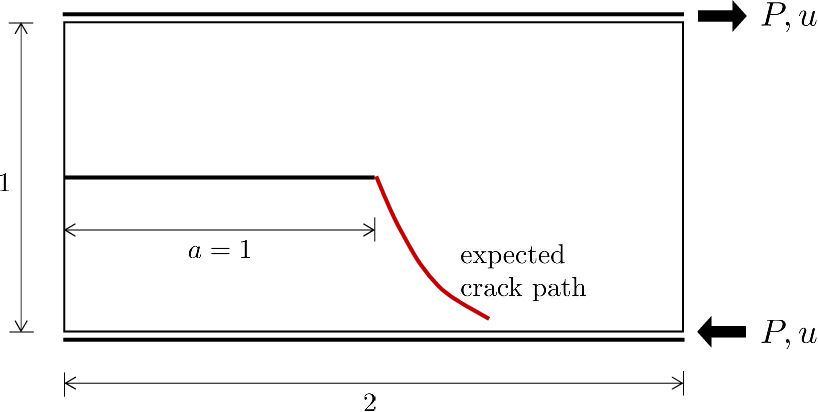}
	\caption{Schematics of the initial specimen geometry and boundary conditions for a pre-notched sample under simple shear. The expected crack path is shown in red.}\label{Fig2}
\end{figure}

In this section, we deploy the phase-field theory (\ref{BVP-u-theory})--(\ref{BVP-v-theory}) to do a numerical study of a rectangular specimen with a pre-existing crack, subjected to simple shear loading. This is considered a benchmark problem to investigate tension-compression asymmetry of phase-field models \cite{Bourdin08, miehe2010}. Fig \ref{Fig2} shows a schematic of the initial geometry of the specimen and applied boundary conditions, as well as the expected crack path normal to the direction of tensile principal stress. The classical phase-field model without an energy split \cite{Bourdin00, Bourdin08} predicts an invalid branching of the pre-existing crack into a tension crack and a compression crack. However, models with energy splits are all able to capture the correct crack path.

Following \cite{miehe2010}, we set Young's modulus $E$ and Poisson's ratio $\nu$ to be $210$ GPa and $0.3$ respectively, while the critical energy release rate $G_c$ is set to $2700$ N/m. The tensile strength $\sts$ is fixed to be $3.26$ GPa and the compressive strength $\scs$ is varied.
With these material parameters, the value of the classical material's characteristic length for the AT$_1$ model is $(3 E \, G_c)/(8 \sts^2) = 0.02$ mm \cite{Tanne18}.
The regularization length $\varepsilon$, which is an independent model parameter, is varied between 0.015 mm and 0.0075 mm. 
% Table 1 provides the values of the regularization length $\varepsilon$, the finite element mesh size $h$, and the corresponding values of the parameter $\de$ for all the results presented below with the Asymptotic Model (\ref{cehat-2020}) and Finite-$\varepsilon$ Model (\ref{cehat-2022}) for the driving force $\ce$.

\begin{figure}[h!]
	\centering
	\includegraphics[width=6.4in]{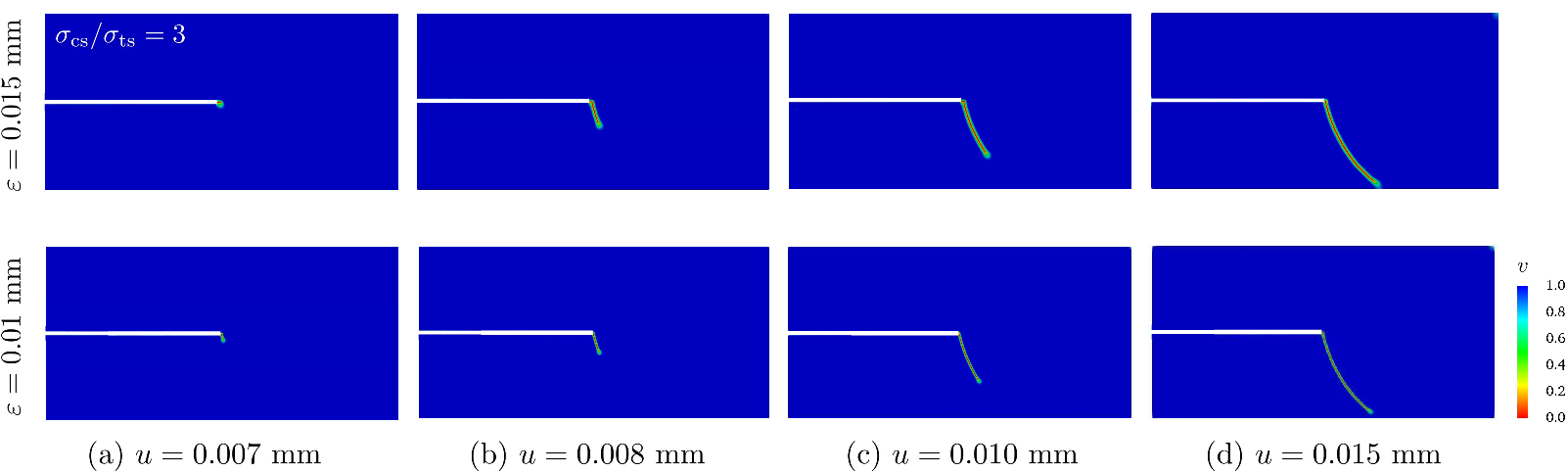}
	\caption{Contour plots of the phase field $v$ predicted by the theory with the Asymptotic Model for driving force $\ce$ for $\scs/\sts=3$ and two values of localization length, $\varepsilon = 0.015, 0.01$ mm. Results are shown for four values of applied displacement: (a) $u=0.007$ mm, (b) $u=0.008$ mm, (c) $u=0.010$ mm, and (d) $u=0.015$ mm. }\label{Fig3}
\end{figure}

Brittle materials typically have a ratio of compressive strength to tensile strength greater than 3. Hence, with the Asymptotic Model, we first investigate the case when $\scs/\sts = 3$.  Figure \ref{Fig3} presents snapshots of the phase field $v$ predicted by the theory at different values of applied displacement for two values of localization lengths $\varepsilon=0.015$ mm and $\varepsilon=0.01$ mm. We observe that the theory predicts the physically expected crack path for both values of $\varepsilon$ and the results are independent of $\varepsilon$. The compression crack is suppressed naturally in this case due to the tension-compression asymmetry built into the theory by accounting for the entire strength surface of the material. Similar results are obtained for all cases with $\scs/\sts > 3$. 
\begin{figure}[h!]
	\centering
	\includegraphics[width=5.5in]{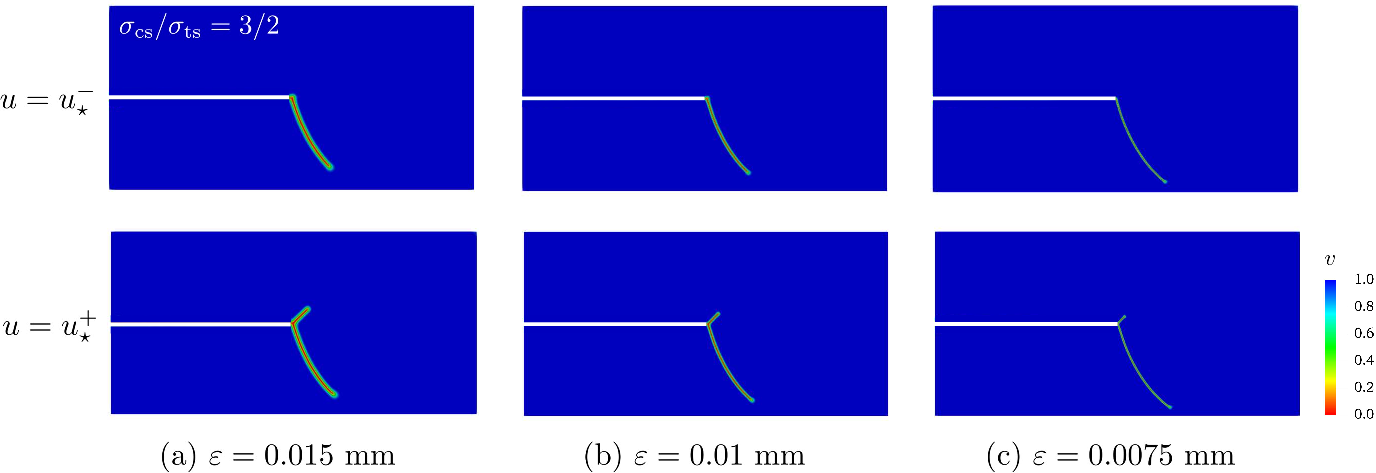}
	\caption{Contour plots of the phase field $v$ predicted by the Asymptotic Model for driving force $\ce$ for $\scs/\sts=3/2$ at displacements just smaller and just larger than the critical displacement, $u_\star$ at which the compression crack nucleates. Results are shown for three localization lengths: (a) $\varepsilon = 0.015$ mm, (b) $\varepsilon = 0.01$ mm, and (c) $\varepsilon = 0.0075$ mm. }\label{Fig4}
\end{figure}
Next, we study a case where compressive strength is of similar value as tensile strength to investigate the limits of the Asymptotic Model. We set $\scs/\sts = 1.5$ and carry out simulations for three values of localization length $\varepsilon=0.015, 0.01$ and $0.0075$ mm. We observe that while initially, only a tension crack nucleates and propagates like the previous case, a compression crack eventually nucleates from the pre-existing crack. Figure \ref{Fig4} shows the contour plots of phase field $v$ for all three values of $\varepsilon$ at two instants: shortly before and shortly after the compression crack nucleates. The compression crack nucleates for all three cases, however, it is observed visually that the tension crack propagates more before the compression crack nucleates as the value of $\varepsilon$ is decreased. Further evidence for this observation is provided by a plot of reaction force $P$ as a function of applied displacement in Figure \ref{Fig5}(a). It is plain from this plot that lowering the value of $\epsilon$ has a substantial impact on the critical applied displacement at which compression crack nucleates. The plot of the implied strength surface for the three values of $\varepsilon$ also shows non-convergence. A prohibitively small value of localization length is likely required to get a converged and physically realistic solution.
\begin{figure}[h!]
	\centering
	\includegraphics[width=5in]{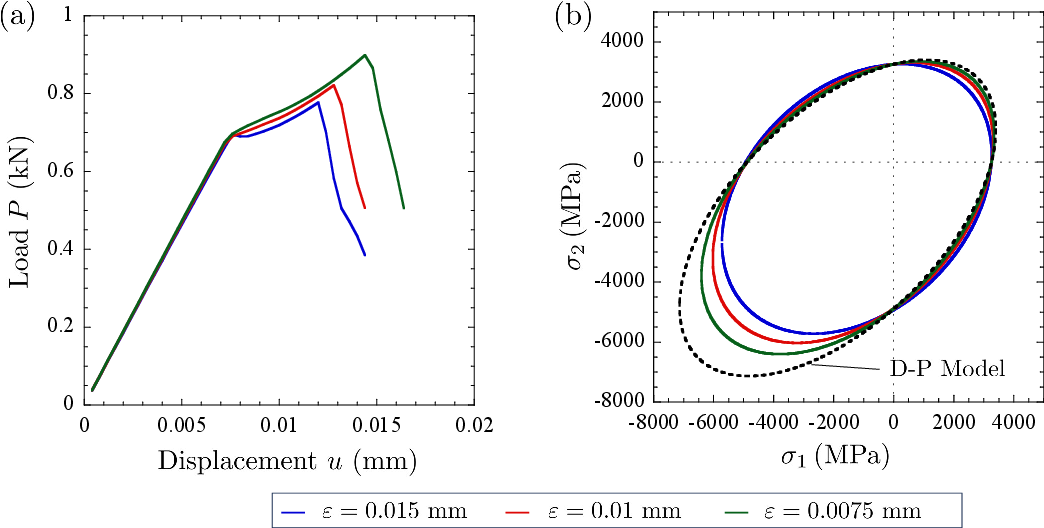}
	\caption{(a) Load-displacement curves and (b) strength surface with the Asymptotic Model for driving force for three values of localization lengths $\varepsilon = 0.015, 0.01$ and $0.0075$ mm.}\label{Fig5}
\end{figure}

\begin{figure}[h!]
	\centering
	\includegraphics[width=5.5in]{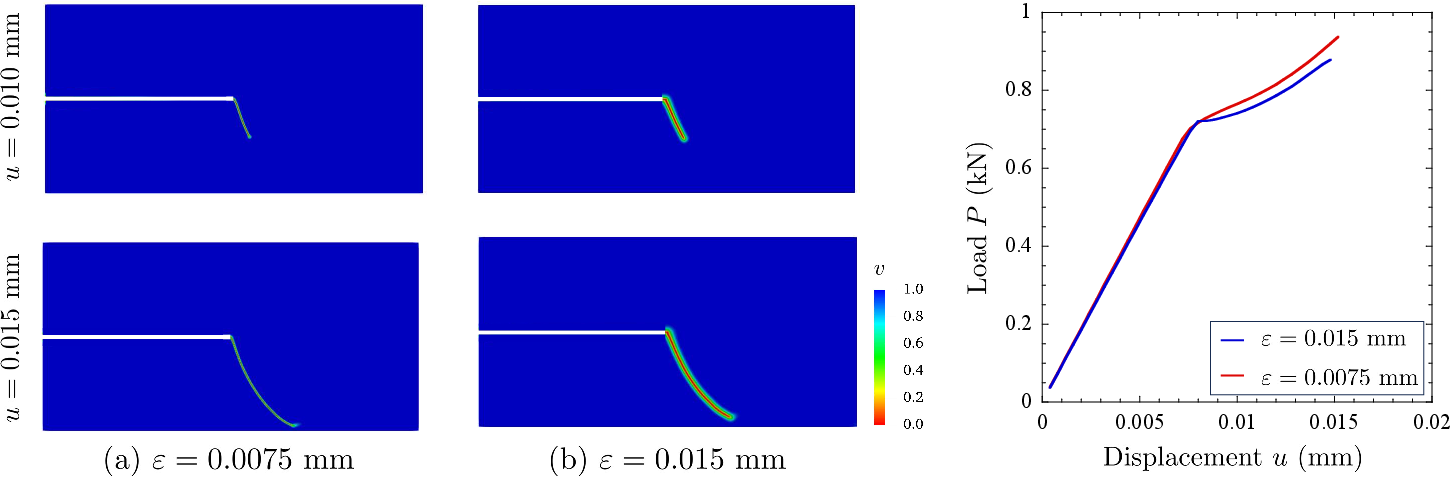}
	\caption{Contour plots of the phase field $v$ predicted by the theory with the Finite-$\varepsilon$ Model for driving force $\ce$ for $\scs/\sts=3/2$ and two values of localization length: (a) $\varepsilon = 0.0075$ mm and (b) $\varepsilon = 0.015$ mm. Load-displacement curves for both values of $\varepsilon$ are shown on the right. }\label{Fig6}
\end{figure}

Our previous results for indentation and Brazilian fracture problems have shown that the Finite-$\varepsilon$ Model for driving force does not suffer from the same limitations. To verify, we carry out the simulations with that model for $\scs/\sts = 1.5$ and the lowest value of localization length, $\varepsilon=0.0075$ mm, adopted in the simulations with the Asymptotic Model. The contour plots for the phase-field $v$ are plotted in Figure \ref{Fig6} (a). No compression crack is observed to nucleate. A simulation is also carried out for a higher value of localization length, $\varepsilon=0.015$ mm, as shown in Figure \ref{Fig6} (b). The absence of a compression crack in that result indicates that the Finite-$\varepsilon$ Model can even be used with larger values of $\varepsilon$. The results with larger values of $\scs/\sts$ with the Finite-$\varepsilon$ Model are not included here, but the same conclusions can be drawn from them.
%, for which the strength surface representation is presumably inadequate, see Remark 1.
%

It is plain from these simulations that while the Asymptotic Model can capture tension-compression asymmetry naturally in problems involving symmetric tension and compression states for realistic values of $\scs/\sts$ for brittle materials, it fails to achieve that for comparable values of $\scs$ and $\sts$ with finite values of $\varepsilon$. The Finite-$\varepsilon$ Model in contrast is more robust and can describe the asymmetry for any values of $\scs/\sts$ and $\varepsilon$. Recall that this model is identical to the Asymptotic Model in the limit $\varepsilon \searrow 0$.

\section{The wing cracks problem} \label{Sec: wingcrack}

\begin{figure}[h!]
	\centering
	\includegraphics[width=5in]{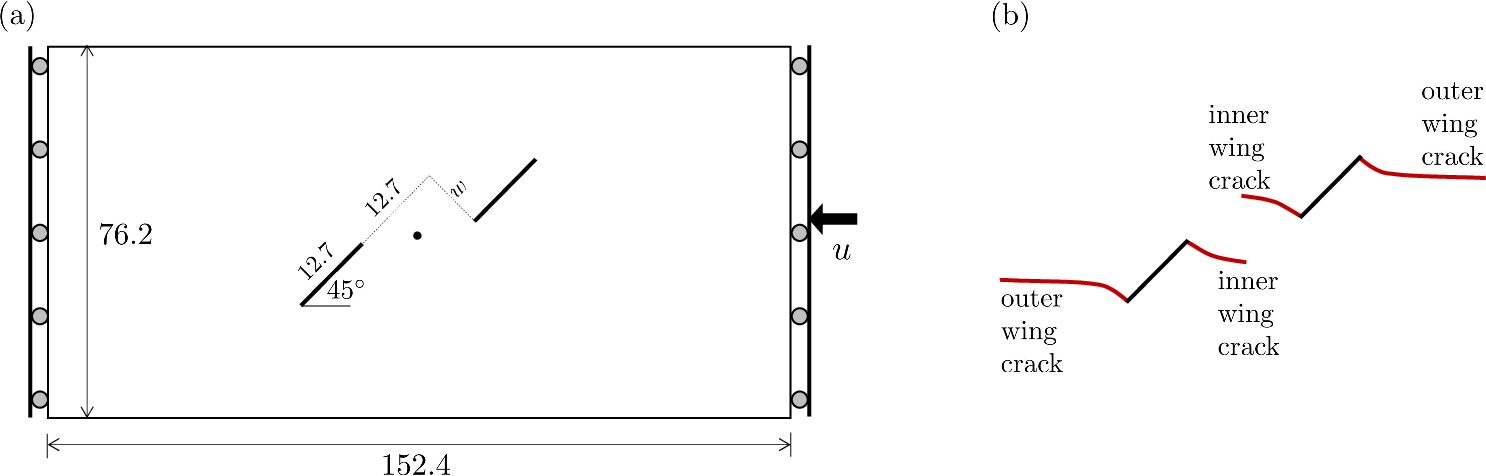}
	\caption{Schematics of the initial specimen geometry and boundary conditions for the wing cracks evolution problem. The expected inner and outer wing crack paths based on experimental observations are shown on the right.}\label{Fig7}
\end{figure}

Next, we confront the theory (\ref{BVP-u-theory})--(\ref{BVP-v-theory}) and the two models for driving force $\ce$ (\ref{cehat-2020}, \ref{cehat-2022}) with a problem of crack nucleation from pre-existing cracks involving large global compressive forces.
Fig \ref{Fig7}(a) shows a schematic of the initial geometry of the specimen and applied boundary conditions. In this popular test for rock-type materials, a rectangular specimen with two (one or more) inclined cracks is subjected to uniaxial compression. Experimental observations \cite{BobetEinstein1998, sagong2002coalescence, lee2011experimental, zhou2021compression} show that the so-called wing cracks (inner and outer) nucleate from existing cracks under this loading; see Fig \ref{Fig7}(b).
Secondary shear cracks are later observed to form. 
The wing crack nucleation has been extensively studied in the literature with two geometries: closed-flaw and open-flaw. For the closed-flaw geometry where the initial crack faces are in contact, several authors have theoretically and experimentally shown that the angle of wing cracks with the pre-existing cracks is significantly affected by the friction between the crack faces \cite{nemat1982compression, horii1985compression, steif1984crack, BobetEinstein1998, germanovich1994mechanisms}. For the open flaw geometry, friction is not as important. In this work, for computational ease, we assume an open flaw geometry which allows us not to consider friction and normal contact between the crack faces in the model. This assumption was also made in previous phase-field studies of this problem \cite{zhang2017wing,BryantSun2018, SteinkeKaliske2022}.

Several authors have also investigated this problem with the classical phase-field method \cite{zhang2017wing, BryantSun2018, zhou2018phase, you2020wing, SteinkeKaliske2022, wick2022} and have noted the inability of the classical energy split techniques, in particular, the two most popular energy split techniques -- volumetric-deviatoric and spectral, to accurately describe the crack paths observed in the experiments. For direct comparison, we first simulated this problem with the classical model and two popular energy splits. We adopt representative values of material parameters for gypsum \cite{SteinkeKaliske2022}. Young's modulus $E$, Poisson's ratio $\nu$, and critical energy release rate $G_c$ are set to $5.96$ GPa, $0.15$ and $10$ N/m, respectively. { The localization length $\varepsilon$ is chosen to be 0.5 mm and mesh size $h$ as $\varepsilon/5$ in all simulations below unless otherwise stated.} The separation between the cracks, $w$, is taken as 6.35 mm.

\begin{figure}[h!]
	\centering
	\includegraphics[width=6.4in]{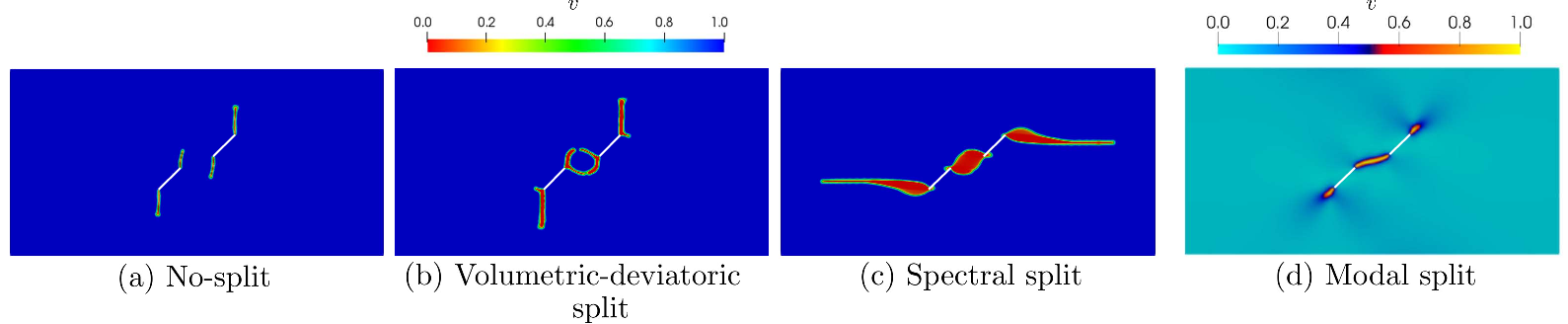}
	\caption{Contour plots of the phase-field $v$ predicted by the (a) classical phase-field (no-split) model, (b) classical phase-field model with volumetric-deviatoric energy split, (c) classical phase-field model with spectral split, and (d) phase-field model with modal split reproduced from Bryant and Sun \cite{BryantSun2018}.}\label{Fig8}
\end{figure}
{ The results are shown in Figure \ref{Fig8}(a)-(c). It is plain from these results these classical energy splits perform poorly for this problem. The volumetric-deviatoric split leads to compression cracks. It also shows an abnormal thickening of the compression crack compared to the no-split case. The results with spectral split show outer wing cracks---but they appear as artificially wider and distorted---and a large damaged region in the center instead of inner wing cracks. The deficiency of spectral split was first noted in Zhang et al. \cite{zhang2017wing}.

\begin{figure}[h!]
	\centering
	\includegraphics[width=6.4in]{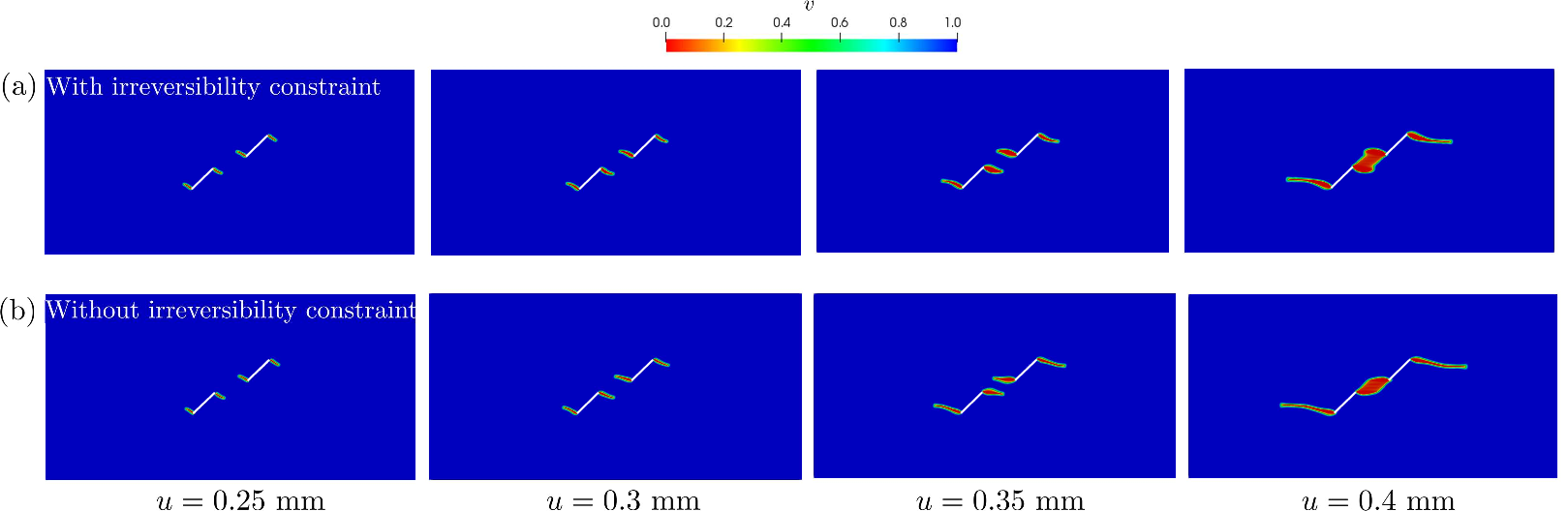}
	\caption{{ Evolution of the phase-field $v$ with applied displacement $u$ for spectral split (a) with irreversibility constraint on $v$, and (b) without irreversibility constraint on $v$.}}\label{Fig-spectral}
\end{figure}
For more insight into the spectral split case, we show in Fig.~\ref{Fig-spectral}(a) the evolution of cracks with the applied displacement. It is observed that the initiation of inner and outer wing cracks is consistent with experiments; however, as these cracks propagate, they tend to get abnormally thicker until the inner cracks coalesce to form a large damaged region. The results suggest an issue with the crack driving force in the spectral split model. The issue can be observed by removing the irreversibility constraint on the phase field. As observed in Fig.~\ref{Fig-spectral}(b), when the phase field is allowed to increase, nucleated cracks change their angle with the pre-existing cracks as they propagate, healing themselves partially and resulting in less thick cracks. Vicentini et al. \cite{LorenzisMaurini2023nucleation} has previously discussed crack thickening due to presence of residual stresses in some models. 
% The recently introduced star-convex model \cite{LorenzisMaurini2023nucleation} as a variation on the volumetric-deviatoric split to account for multiaxial crack nucleation also shows unphysical behavior.
The deficiencies of the classical splits under large compressive forces have also been noticed before for the problem of indentation of glass plates with flat-ended indenters \cite{StroblSeelig, KRLP22}.  

Alternative energy splits have been proposed to resolve these deficiencies. One such split is the modal split of \cite{BryantSun2018}, in which the energy is first split based on the different kinematic modes -- opening mode and shearing mode. The opening mode energy is taken to be non-zero only when a scalar mode I strain is greater than 0. However, this energy split predicts a primary shear crack instead of wing cracks when the fracture toughness is mode-independent, see Fig.~\ref{Fig8}(d), where a result reproduced from \cite{BryantSun2018} is shown. {  The so-called directional splits \cite{steinke2019, SteinkeKaliske2022}  also show similar behavior. Based on the results with these splits, it has been claimed that primary wing cracks can only be obtained when the mode II fracture toughness is significantly higher than the mode I fracture toughness.}

}

To simulate this problem with the theory of Kumar et al. (\ref{BVP-u-theory})--(\ref{BVP-v-theory}) and the two models for driving force $\ce$, we fix tensile strength $\sts$ of gypsum to $3$ MPa and we choose the compressive strength as 20 MPa or 40 MPa. The fracture toughness is considered mode-independent.
The parameter $\de$ can be calculated with the equation (\ref{delta-eps-final-h}).
% Table 2 provides the values of the regularization length $\varepsilon$, the finite element mesh size $h$, and the corresponding values of the parameter $\de$ for the results with the theory of Kumar et al. 
We model the pre-existing cracks as phase-field cracks by setting phase-field $v$ equal to 0 on the crack surfaces. Modeling the cracks as open flaws with a small distance between the crack surfaces yields similar results as shown later below.

\begin{figure}[h!]
	\centering
	\includegraphics[width=5in]{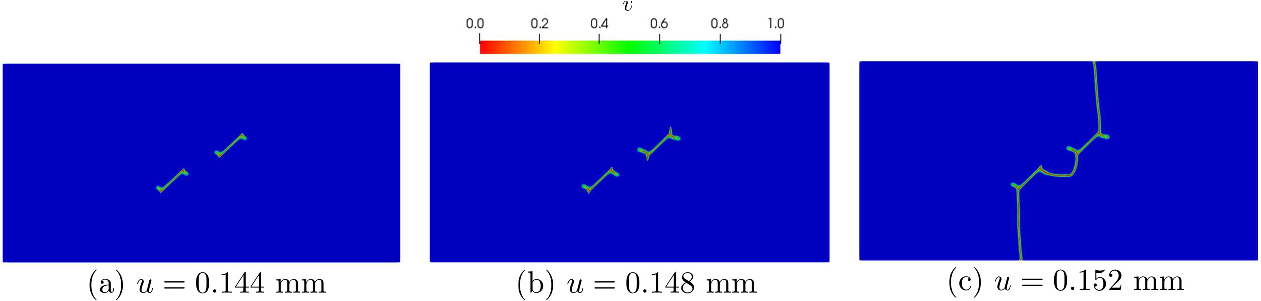}
	\caption{Contour plots of the phase field $v$ predicted by the theory with the Asymptotic Model for driving force $\ce$ for wing crack problem with $w=6.35$ mm and  $\scs/\sts=40/3$ }\label{Fig9}
\end{figure}
Figure \ref{Fig9} presents results with the Asymptotic Model for $\ce$ with $w=6.35$ mm. Fig.~\ref{Fig9}(a) shows that both inner and outer wing cracks initially nucleate in the specimen. However, soon after, compressive cracks nucleate and propagate fast toward the lateral boundary as seen in Fig.~\ref{Fig9}(b)-(c). Unlike the simple shear example in Section 3, a larger value of compressive strength does not suppress the nucleation of compressive cracks, however it does delay it.

\begin{figure}[h!]
	\centering
	\includegraphics[width=6.4in]{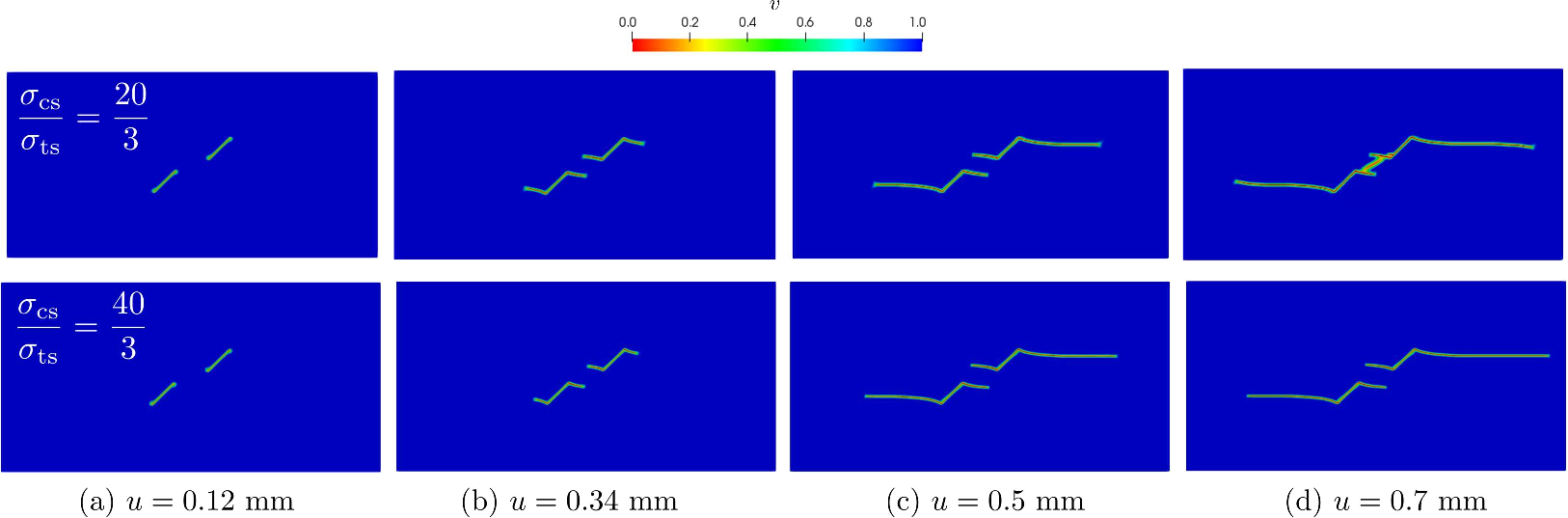}
	\caption{Contour plots of the phase field $v$ predicted by the theory with the Finite-$\varepsilon$ Model for driving force $\ce$ for wing crack problem $w=6.35$ mm and two values for compressive strength to tensile strength ratio: (a) $\scs/\sts=20/3$ and (b) $\scs/\sts=40/3$, at four applied displacements $u$.}\label{Fig10}
\end{figure}
\begin{figure}[h!]
	\centering
	\includegraphics[width=3.5in]{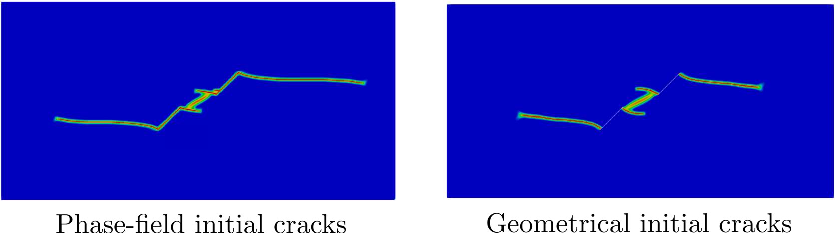}
	\caption{{ {  Comparison for the predicted wing crack and shear crack formation between domain with initial phase-field cracks and domain with initial geometrical cracks with a small separation between crack faces. }} }\label{Fig-shear}
\end{figure}

Figure \ref{Fig10}(a) presents results with the Finite-$\varepsilon$ Model for $\ce$ with $\scs/\sts=20/3$. The contour plots of the phase field $v$ are shown at four values of applied displacements $u$.
Two comments are in order for these results. First, the Finite-$\varepsilon$ Model for the driving force is able to capture the experimentally observed inner and outer wing crack paths. We observe that the inner and outer wing cracks nucleate at an applied displacement of $u=0.12$ mm and stably propagate in the direction of applied loading.  Second, the model can also capture the formation of secondary shear cracks and the coalescence of initial cracks. At around $u=0.7$ mm, a secondary shear crack forms in between the inner wing cracks. However, the coalescence is a brutal crack event that may also cause violation of material impenetrability, and we experience numerical convergence issues beyond this displacement. { The results with geometrical initial open flaws with a small spacing between the crack surfaces show similar results; however, with the shear crack forming slightly earlier, as shown in Fig.~\ref{Fig-shear}.}

Fig.~\ref{Fig10}(b) shows the results obtained for $\scs/\sts=40/3$. The results are qualitatively and quantitatively similar to the case with $\scs/\sts=20/3$, except that because of the higher compressive strength and consequently higher shear strength, the secondary shear crack nucleates later. {  The shear crack develops in a manner similar to the previous case, but local violations of material impenetrability constraint hinder a fully converged solution. Accepting a higher tolerance for convergence, we can observe the emergence of external shear cracks and secondary tensile cracks. A more thorough experimental comparison will likely require proper modeling the material impenetrability constraint.}

{ 
To better understand why the Finite-$\varepsilon$ model can capture tension-compression asymmetry, while the Asymptotic model (for finite $\varepsilon$) and energy splits models cannot, we perform a purely elastic analysis of this problem, tracking the pointwise violation of the respective strength surface $\hat{\mathcal{F}}(\boldsymbol{\sigma})$ predicted from each model. For the classical variational models with splits, the strength surface is defined by the algebraic equation
\begin{equation}
    \hat{\mathcal{F}}(\boldsymbol{\sigma})= 2 W^{+} (\boldsymbol{\sigma}) - \dfrac{3 G_c}{8 \varepsilon}=0
\end{equation}
where $W^{+}$ represents the `tensile' part of the strain energy. For both the asymptotic and Finite-$\varepsilon$ model, $\hat{\mathcal{F}}(\boldsymbol{\sigma})=0$ is defined as the right-hand side of the equation (\ref{BVP-v-theory})$_1$ equal to 0 with $v=1$. The violation of the strength surface is a necessary condition for a crack nucleation. Fig.~\ref{Fig-strength} presents contour plots of the regions (shown in black) of the specimen near one of the inclined cracks at which the strength surface predicted by the different models is met. 
% The results are shown for the classical AT$_1$ variational model without energy split and with volumetric-deviatoric and spectral splits, as well as for the asymptotic and Finite-$\varepsilon$ models. 
Strength violations occur in both tensile and compressive regions for the no-split, volumetric-deviatoric split, and the Asymptotic model. In contrast, the spectral split and Finite-$\varepsilon$ models show strength violations only in the tensile region. Since the full phase-field fracture analysis revealed that the no-split, volumetric-deviatoric split, and Asymptotic models lead to spurious compression cracks, while the spectral split and Finite-$\varepsilon$ models do not, these elastic analysis results strongly support the idea that correctly incorporating the material's strength surface can naturally account for tension-compression asymmetry in crack propagation.
\begin{figure}[h!]
	\centering
	\includegraphics[width=6.4in]{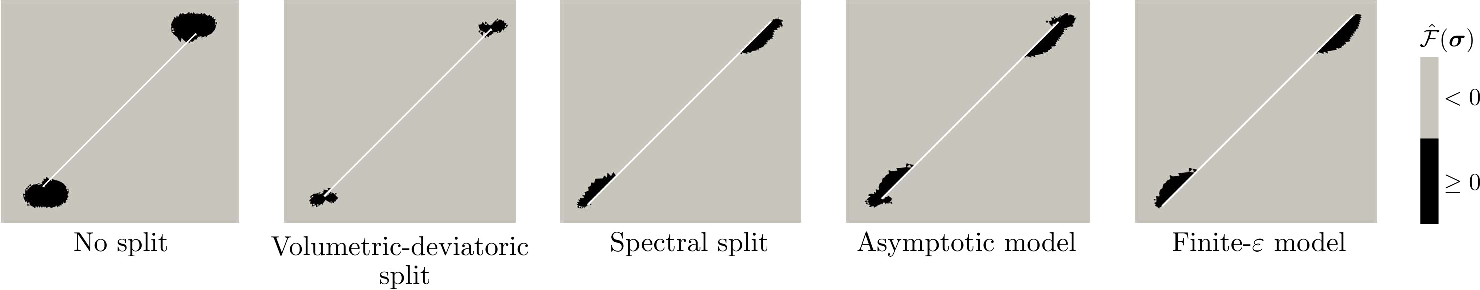}
	\caption{{ {  Contour plots of the regions of the specimen near one of the inclined cracks where the elastic stress field exceeds the strength surface predicted by different phase-field models.}} }\label{Fig-strength}
\end{figure}

}

\begin{figure}[h!]
	\centering
	\includegraphics[width=5in]{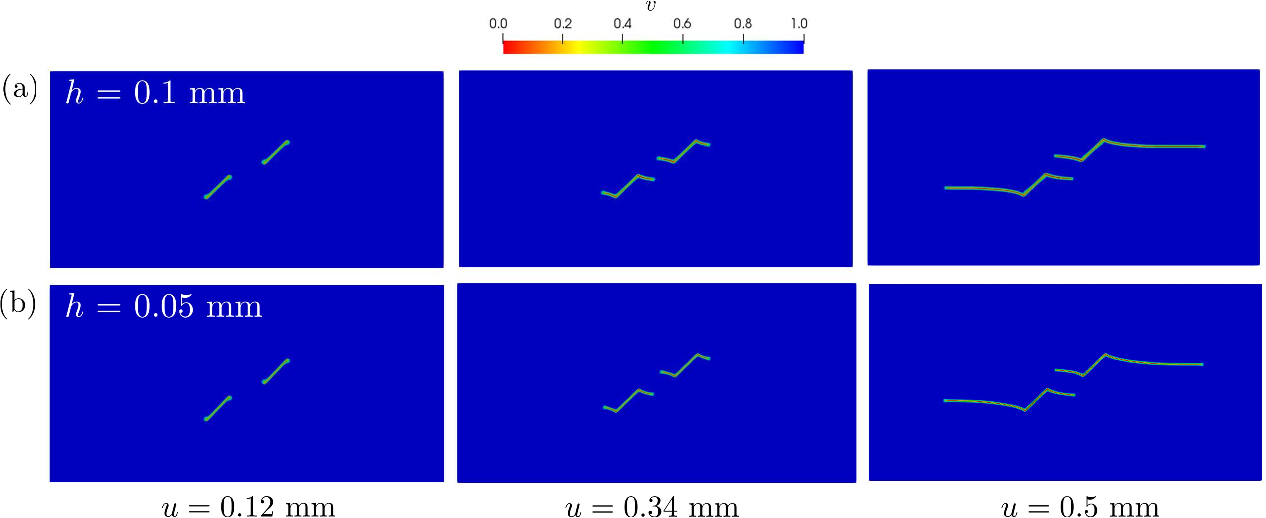}
	\caption{{ Convergence of results for phase-field evolution with mesh size for fixed regularization length $\eps=0.5$ mm. Two mesh sizes are studied: (a) $h=0.1$ mm, and (b) $h=0.05$ mm.} }\label{Fig-meshsize}
\end{figure}
\begin{figure}[h!]
	\centering
	\includegraphics[width=5in]{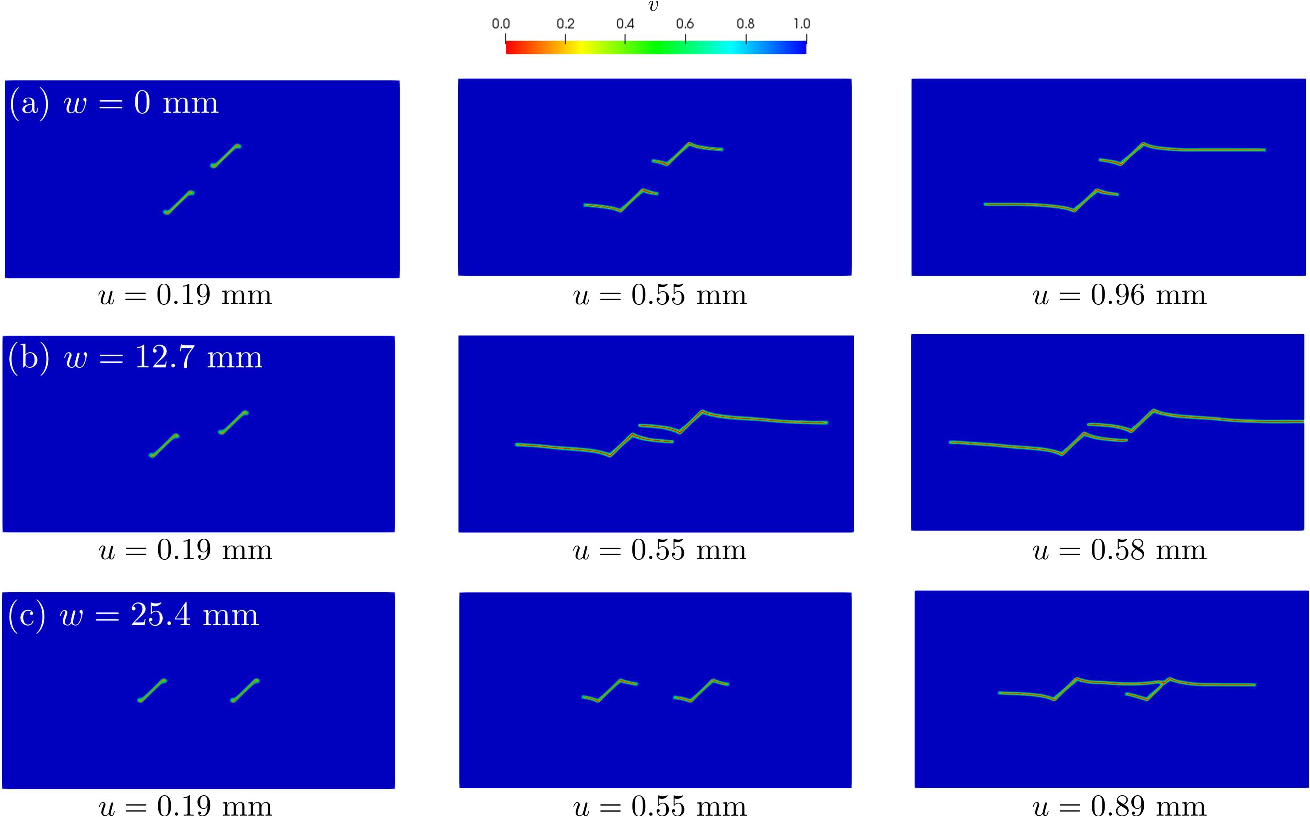}
	\caption{Contour plots of the phase field $v$ for wing crack problem with three different values of separation between the two pre-existing cracks -- (a) $w=0$ mm, (b) $w=12.7$ mm, and (c) $w=25.4$ mm --  predicted with the Finite-$\varepsilon$ Model for driving force $\ce$.}\label{Fig11}
\end{figure}

{ For completeness, we carried out the same simulation with a two times uniformly refined mesh to show that the results are converged with respect to mesh size. Fig.~\ref{Fig-meshsize} shows the evolution of phase-field with displacement, which is practically the same for the two uniform mesh sizes, $h=0.1$ mm and $h=0.05$ mm, studied.} Furthermore, we also carried out simulations for three different separations between the two pre-existing cracks as is commonly done in experimental studies shown in Fig.~\ref{Fig11}. In each case, we observe inner and outer wing cracks nucleating at roughly the same angle as the base case with $w= 6.35$ mm and propagating in the loading direction towards the center of the left and right boundaries. 

\section{Final Comments}\label{Sec: Final Comments}

{
The two examples presented in this paper demonstrate the capability of the theory of Kumar et al. (2018, 2020) for linear elastic brittle fracture to show tension-compression asymmetry in crack propagation. As such, it provides a more natural way to model fracture under compression than the traditional energy split approach.
The examples also highlight the differences between the results from the two models for the driving force that describes the material's multiaxial strength surface. The first model proposed in Kumar et al. (2020) successfully suppresses compression cracks for realistic brittle materials when the problem involves a symmetric tension-compression state, like the simple shear problem in Section 3. 
However, for localization lengths $\varepsilon$  that are not prohibitively small, it fails in problems involving large compressive forces. 
The second model proposed in Kumar et al. (2022) to improve the representation of strength surface for finite values of localization length under compressive states proves to be more robust and reliably provides physically consistent results for a wide range of values for $\varepsilon$ and compressive strength to tensile strength ratio.

In retrospect, indirect evidence of tension-compression asymmetry can also be found in the recent studies on indentation test \cite{KRLP22} and Brazilian fracture test \cite{KLDLP23}, which both involve punching loads. While these studies are primarily about crack nucleation under multiaxial stress states in the absence of pre-existing cracks, the presence of tension-compression asymmetry in the model should be necessary to get the experimentally consistent propagation of the nucleated cracks. While direct analysis has not been presented on these problems, the results of this work provide some confidence about the ability of the model to describe the tension-compression asymmetry  in crack propagation under punching loads.

% add to the validation results presented in previous works on the indentation problem \cite{KRLP22} and the Brazilian fracture test \cite{KLDLP23} which demonstrate that the theory of Kumar et al. (2020) for linear elastic brittle fracture can describe the tension-compression asymmetry in fracture nucleation and propagation under a wide variety of loading and geometrical situations. 

The results presented in Section 4 for the wing cracks problem have also made it plain that a complete model of tension-compression asymmetry with the phase-field models is needed to consistently and uniquely interpret the fracture toughness from experimental results involving large compression and mixed mode fracture. Specifically, they have shown that wing cracks can be obtained even when the material has a mode-independent fracture toughness. Future work will investigate the three-dimensional effects in this problem. The 3-D peculiarities of crack growth in rock under compression have been noted in the literature \cite{germanovich1993, germanovich1994mechanisms} and can result in dynamic, burst-like failure instead of the stable crack growth seen in 2-D.

% From a computational standpoint, we will investigate in the future if the governing equations in the theory of Kumar et al. can be solved monolithically. 
The search for a complete phase-field model of fracture nucleation and propagation for any material, brittle or ductile, under any loading conditions, static or dynamic, is ongoing. The results of this work add to the evidence accumulated over the last several years that the phase-field theory of Kumar et al. may have the necessary ingredients for, at least, brittle materials, although alternative formulations with the same ingredients are certainly possible.} {  Finally, accounting for the material impenetrability and the proper transfer of stresses in self-contact remains an issue with all phase-field formulations. }

% Overall, the results of this paper add to the other qualitative and quantitative validation results presented in previous work that have demonstrated the same PDEs is able to describe fracture nucleation and propagation in any brittle material, soft or hard, for any quasi-static monotonic loading conditions

\section*{Acknowledgements}

\noindent The author A. Kumar would like to acknowledge the financial support from the start-up fund provided by the Georgia Institute of Technology.

\bibliographystyle{elsarticle-num-names}
\bibliography{ref}

\end{document}